\documentclass[a4paper,preprintnumbers,floatfix,superscriptaddress,pra,twocolumn,showpacs,notitlepage,longbibliography]{revtex4-1}
\usepackage[utf8]{inputenc}
\usepackage[T1]{fontenc}
\usepackage[sc,osf]{mathpazo}
\usepackage{amsmath, amsthm, amssymb,amsfonts,mathbbol,amstext}
\usepackage{graphicx}
\usepackage{dcolumn}
\usepackage{bm}
\usepackage{hyperref}
\usepackage{mathtools}
\usepackage{comment}
\usepackage{color}
\usepackage{multirow}
\usepackage{relsize}

\DeclareUnicodeCharacter{2009}{-}

\def\1{\mathbf{1}}
\def\0{\mathbf{0}}

\newcommand{\ket}[1]{| #1 \rangle}

\newcommand{\mean}[1]{\left\langle #1 \right\rangle}

\renewcommand{\rho}{\varrho}

\newcommand{\processnext}[1]{%
  \ifx\listfinish#1\empty\else\listact{#1}\expandafter\processnext\fi}

\definecolor{airforceblue}{rgb}{0.36, 0.54, 0.66}
\definecolor{gabirgb}{rgb}{0., 0.7, 0.2}

{\begin{framed}\begin{small}}
{\end{small}\end{framed}}

\DeclareGraphicsExtensions{.pdf,.png,.jpg}

\makeindex

\begin{document}
\title{Causal modeling the delayed choice experiment}
\date{\today}

\author{Rafael Chaves}
\author{Gabriela Barreto Lemos}
\author{Jacques Pienaar}
\affiliation{International Institute of Physics, Federal University of Rio Grande do Norte, 59078-970, P. O. Box 1613, Natal, Brazil}

\begin{abstract}
Wave-particle duality has become one of the flagships of quantum mechanics. This counter-intuitive concept is highlighted in a delayed choice experiment, where the experimental setup that reveals either the particle or wave nature of a quantum system is decided after the system has entered the apparatus. Here we consider delayed choice experiments from the perspective of device-independent causal models and show their equivalence to a prepare-and-measure scenario. Within this framework, we consider Wheeler's original proposal and its variant using a quantum control and show that a simple classical causal model is capable of reproducing the quantum mechanical predictions. Nonetheless, among other results, we show that in a slight variant of Wheeler's Gedankenexperiment, a photon in an interferometer can indeed generate statistics incompatible with any non-retrocausal hidden variable model whose dimensionality is the same as that of the quantum system it is supposed to mimic. Our proposal tolerates arbitrary losses and inefficiencies making it specially suited to loophole-free experimental implementations.

\end{abstract}

\maketitle

\begin{figure}[t]
\centering\includegraphics[width=0.8\linewidth]{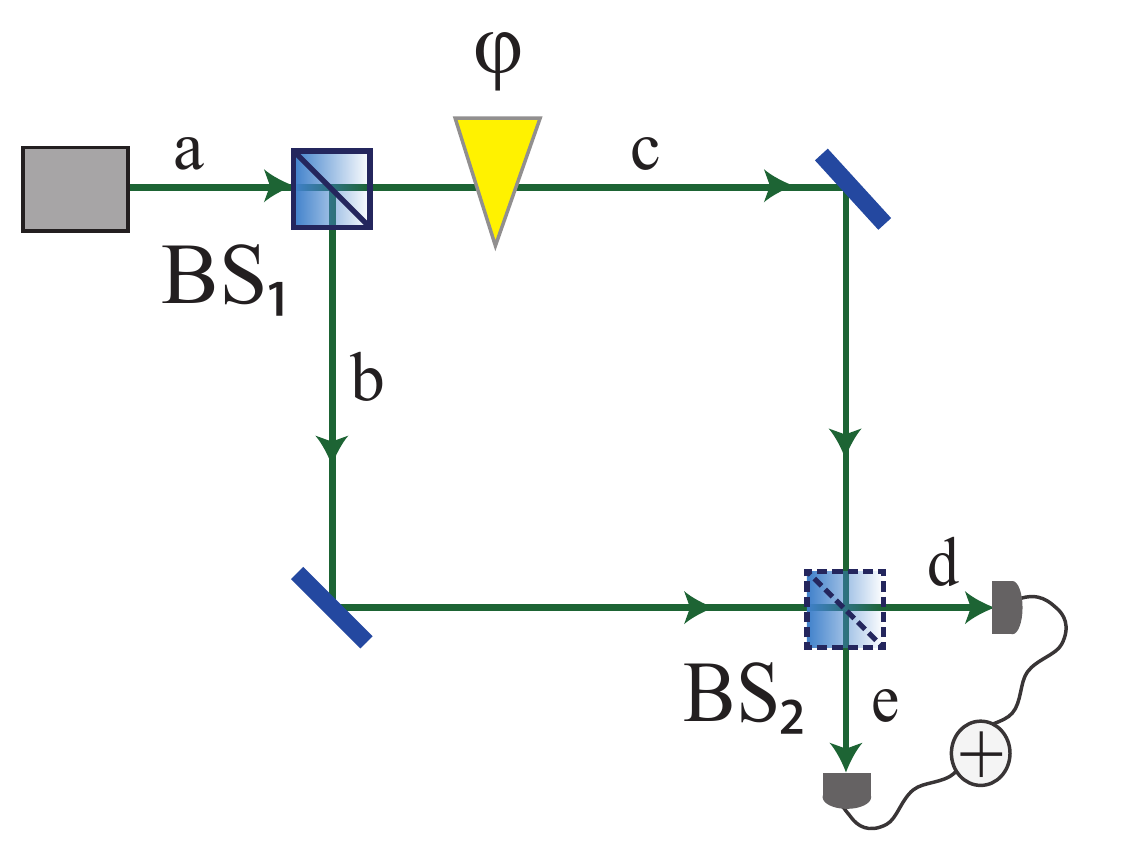}
\caption{In Wheeler's delayed choice \textit{Gedankenexperiment}, the choice of removing or not the beam-splitter $\mathrm{BS}_2$ in a Mach-Zehnder interferometer is made after the photon has entered the interferometer (at $\mathrm{BS}_1$). With $\mathrm{BS}_2$ present, the photon counting rate at either detector is a function of $\phi$; when absent, the counting rate is independent of $\phi$.}
\label{WheelerDC}
\end{figure}

Wave-particle duality is at the heart of the most renowned debates in quantum theory. Although light and matter produce individual counts on a detector, they also exhibit interference in certain experimental arrangements. In Wheeler's  delayed choice experiment (WDCE) (Fig.\ref{WheelerDC}) \cite{Wheeler1978,Wheeler1984}, the
experimenter chooses whether or not to remove the beam-splitter $\mathrm{BS}_2$ \emph{after} a photon has entered a Mach-Zehnder interferometer (at $\mathrm{BS}_1$), thereby observing no-interference (particle-like behavior) or interference (wave-like behavior) accordingly \cite{note1}. By excluding any causal link between the experimental setup and a hidden variable that predefines the photon's behavior (retrocausality), delayed choice experiments are usually about defining and testing ``wave-particle objectivity'' models, in which a quantum system is intrinsically \textit{either} a wave \textit{or} a particle  (see \cite{Ma2016} and references therein).

Within this mindset, Ionicioiu and Terno suggested a particular wave-particle objective model \cite{Ionicioiu2011} (hereafter called the IT model), which they ruled out using a \textit{quantum delayed choice} experiment (QDCE), realized in \cite{Auccaise2012}, in which the $\mathrm{BS}_2$ in WDCE is replaced by a quantum control that can be in a superposition of being present or absent until after the photon is detected. This reasoning relies upon a \textit{device-dependent} argument that the beam-splitter was truly in a quantum superposition, since the statistics alone could not distinguish a superposition from an incoherent mixture. This motivated \textit{entanglement assisted} QDCE \cite{Ionicioiu2014,Rossi2017,Rab2017}, in particular the experiments in \cite{Kaiser2012,Peruzzo2012}, which rely on the violation of a Bell inequality to rule out the IT model in a \textit{device-independent} (DI) manner, i.e. from the measurement statistics alone without prior assumptions about the quantum nature of the control (\ref{entqdce}). However, the IT model of wave-particle objectivity makes rather strong assumptions that, as we show below, are trivially inconsistent with the assumption of no-retrocausality and the quantum predictions. Could wave-particle objectivity be tested more generally, using causal assumptions rather than specifically tuned models? Furthermore, can we devise a DI proof of the non-classical behaviour of a delayed choice experiment without the need for entanglement (or the violation of a Bell inequality)?
\par
\begin{figure*}
\setlength\fboxsep{0pt}
\setlength\fboxrule{0.25pt}
\centering\includegraphics[width=1\linewidth]{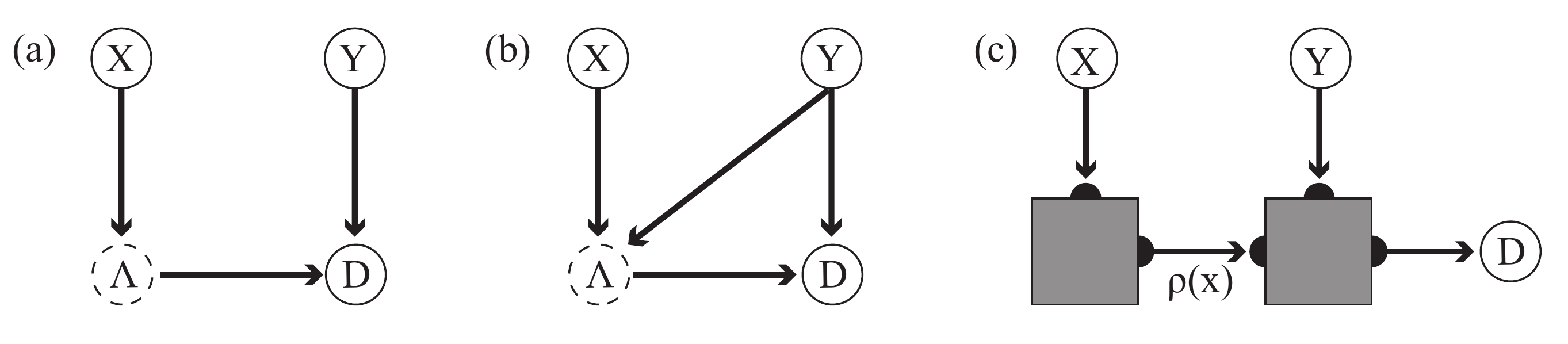}
\caption{(a) The DAG representing the causal model for WDCE.
(b) A causal model allowing for retrocausal influence from $Y$ to $\Lambda$.
(c) The representation of the PAM scenario in terms of preparation and measurement devices (black-boxes).
}
\label{DAGfigures}
\end{figure*}
\par

In this letter we employ tools from the field of causal inference \cite{Pearl2009} to give a positive answer to these questions. We propose that WDCE can be described by a DI ``prepare-and-measure" scenario \cite{Gallego2010}. In this framework, the only relevant constraint on the classical model being tested is its dimension, which leads us to suggest replacing wave-particle objectivity by the assumption that the HV has the same dimension as the quantum system under test. We demonstrate that -- contrary to intuition -- a two-dimensional classical variable \textit{can} explain the outcomes of WDCE (and of the QDCE). We then propose a delayed choice experiment where, instead of removing $\mathrm{BS}_2$, the experimenter can chose to slightly displace it, imparting a relative phase shift on the beam. Using previously derived dimension witnesses inequalities \cite{Gallego2010,Bowles2014}, we can exclude \emph{any} model where the classical variable has at most two dimensions (assuming, e.g, the values ``wave'' and ``particle''). Since we only detect one mode (we do not consider coincidences), our proposal tests the non-classicality of this system without needing to introduce entanglement. Our proposal has the additional advantage of being robust to arbitrarily small (but larger than zero) detection efficiencies and to loss inside the interferometer, making it specially suited for loophole-free experimental implementations. Finally, we quantify how much retrocausal influence would be needed to explain the observations. 

\emph{The delayed-choice experiment from a causal perspective --} 
How does quantum behaviour manifest in the WDCE? On one hand, a single photon in an interferometer can in principle exhibit spatial mode entanglement. On the other hand, Wheeler's argument did not appeal to spatial separation, but to temporal causality: the state leaving the interferometer exists in the causal future of the state it possessed upon entry. This leads us to ask: could quantum behaviour manifest itself even between the \textit{temporal} parts of the WDCE, i.e. between the preparation of its `causes', and the measurement of its `effects'?
 
We begin by reviewing some basic concepts in causal modeling \cite{Pearl2009}. We use uppercase characters to denote random variables, and lowercase to denote their possible values. Given a set of $n$ experimental random variables $X_1,\dots,X_n$, a hypothesis about the causal relationships between these variables can be represented by a directed acyclic graph (DAG) with the variables as nodes. Each arrow represents a direct causal influence of one variable upon another, in the following sense: if $\mathrm{PA}(X_i)$ are the parents (direct causes) of $X_i$ in the DAG, then there exists a local noise variable $U_i$ (having no causes) and a deterministic function $f_i$ such that $x_i=f_i(\mathrm{PA}(x_i),u_i)$ \cite{note3}. Consequently, the joint probability distribution factorizes as a product: $p(\vec{x})=\prod_{i=1}^{n} p(x_i\vert \mathrm{PA}(x_i))$, relative to the DAG. The factorization of probabilities is a DI constraint, as it depends only on the causal relationships described in the DAG and not on the particular choices of functions and noise variables $\{f_i, u_i\}$. This simplifies the task of causal hypothesis testing: if the observed probability does not factorize according to the structure of the hypothesized DAG, we can exclude any causal model based on that DAG from being a valid explanation for our observation. 

We propose that the causal relations in WDCE (Fig. \ref{WheelerDC}) can be represented by the DAG shown in Fig. \ref{DAGfigures}(a). The variable $X$ determines the phase shift $\phi_x$ between the interferometer arms.
The variable $\Lambda$ corresponds to the intrinsic state of the photon upon entering the interferometer, just after the phase $\phi_x$. The values of $\Lambda$ could correspond to possible quantum states, or to values of some hidden variable encoding instructions for its future behavior (In Ref.\cite{Ionicioiu2011}, for example, this would correspond to `particle' or `wave'). For our purposes, only the dimensionality of $\Lambda$ is relevant, its interpretation is unimportant. Since the experiment involves a single photon with two modes, which encodes only a single classical bit \cite{Holevo}, $\Lambda$ should likewise be binary. That is, we propose that the hidden variable (HV) should not be able to encode more classical bits than the quantum system it is supposed to simulate. Assuming no retrocausal influences, the probability of $\Lambda$ respects $p(\lambda \vert x,y)=p(\lambda \vert x)$, depending only on $X$ (plus, possibly, local noise). In particular, it cannot depend on the delayed choice $Y$, which determines the experimental arrangement: $y=1$ when $\mathrm{BS}_2$ is present and $y=0$ when it is removed \cite{note4}. Finally, the variables $D$ and $E$ represent the photon detectors and take the values $d,e \in \{ 1,0 \}$ depending whether the detector has clicked or not, respectively \cite{notexy}.

We will restrict our attention to the probability for the detector $D$ to click \cite{note5}. The causal model in Fig. \ref{DAGfigures}(a) implies that any observed distribution compatible with it should factorize as:
\begin{equation}
\label{HVmodel}
p(d \vert x,y) = \sum_{\lambda} p(d\vert y,\lambda) p(\lambda \vert x) \, .
\end{equation}
For comparison, let us compute the probabilities predicted by quantum mechanics in this setup. We treat the photon in the Mach-Zehnder as a two-level quantum system in the path degree of freedom. The initial state is a single photon in mode $a$. The state emerging from the interferometer is denoted $\ket{\Psi(x,y)}$. In the Fock basis $\ket{\hat{n}_d \hat{n}_e}$ at the output modes $d,e$, it has the form $\ket{\Psi(x,0)} = \frac{1}{\sqrt{2}}\left( \ket{01}+ e^{i\phi_x} \ket{10} \right)$ when $BS_2$ is absent and $\ket{\Psi(x,1)}=\cos(\frac{\phi_x }{2}) \ket{01}-i\sin(\frac{\phi_x }{2})\ket{10}$ when $BS_2$ is present. The probabilities $p(d|x,y)$ for detector $D$ to click are therefore given by $p(d \vert x,0)=1/2$ and $p(1 \vert x,1)=1-p(0 \vert x,1)=\sin^{2}(\frac{\phi_x}{2})$ respectively.

In light of this formulation, we now re-examine arguments put forward in Refs. \cite{Ionicioiu2011,Ionicioiu2015} claiming that no hidden variable model of the form \eqref{HVmodel} could account for the quantum predictions of the QDCE. In that reference, it is assumed that wave-particle objectivity implies $\lambda=\textrm{'wave'}$ if and only if interference is observed, and $\lambda=\textrm{'particle'}$ otherwise\cite{Ionicioiu2011,Ionicioiu2015}. In the present notation, this assumption would require $\Lambda$ to be perfectly correlated with the variable $D$. However, quantum mechanics predicts correlations between $D$ and the delayed-choice variable $Y$, and since $D$ is binary (the detector can either click or not in each run), this leads to the conclusion that $\Lambda$ is correlated with $Y$ -- but this can only be explained by a retrocausal influence of $Y$ on $\Lambda$, which has been excluded by assumption. 
Our causal analysis shows that the assumptions in the IT model are trivially mutually inconsistent.

We now show that, absent any special assumptions about the relationship of $\Lambda$ to the detector response $D$, we can reproduce the predictions of WDCE using a classical two-valued hidden variable. We introduce a local noise term $U_{D}$ for the detector $D$ such that $p(d \vert y,\lambda)=\sum_{u_D}p(d \vert y,\lambda,u_{D})p(u_D)$. Choosing
\begin{eqnarray}
& & i) \quad p(u_D=0)=p(u_D=1)=1/2, \\ \nonumber
& & ii) \quad p(\lambda=0 \vert x)=1- p(\lambda=1 \vert x)= \cos^2{\frac{\phi_x }{2}}\\ \nonumber
& & iii)  \quad p(d \vert y=0,\lambda,u_D)=p(d \vert y=0,u_D)=\delta_{d,u_D}, \\ \nonumber
& & iv) \quad p(d \vert y=1,\lambda,u_D)=p(d \vert y=1,\lambda)=\delta_{d,\lambda},
\end{eqnarray}
one can verify that we recover the predictions of quantum theory for any choices of the phases $\phi_x$. Therefore WDCE can be explained by a classical causal model without the need of retrocausality. In the Appendix we show how this classical model can be extended to the QDCE \cite{Ionicioiu2011}. 

\emph{The delayed-choice as a prepare-and-measure scenario --} Given that there is a causal model (hence a hidden variable model) explaining WDCE, it is natural to ask whether small modifications to the experiment would allow us to rule it out. We will continue to assume no retrocausality and that $\Lambda$ is two-dimensional.
First, we draw a correspondence to the device-independent \textit{prepare-and-measure} (PAM) scenario \cite{Gallego2010}, shown schematically in Fig.\ref{DAGfigures}(c). In the PAM scenario an initial black-box prepares different physical systems (upon pressing a button labeled by $x$) that are then sent to a second black-box where the systems are measured (upon pressing a button labeled by $y$) and produce an outcome labeled by $d$. The essential feature of the PAM scenario is that quantum systems can produce statistical distributions that can only be reproduced by classical systems of higher dimensionality. In particular, there is a quadratic gap between classical and quantum dimensions, as one can devise  situations where the statistics produced by a $\sqrt{k}+1$ dimensional quantum system can only be reproduced by a classical system of at least $k+1$ dimensions \cite{Bowles2014}. Moreover, the PAM scenario implies constraints on the probabilities \cite{Chaves2015entropy} that are equivalent to the causal constraints of the delayed choice experiment (eq. \eqref{HVmodel}). Therefore general results pertaining to the PAM scenario can be readily adapted to analyze the experiment. For instance, for $2k$ preparations (choices of $X$) and $k$ possible measurements (choices of $Y$) the ($k \times k$) matrix \cite{Bowles2014}
\begin{equation}
W_k(i,j)= p(2j,i)-p(2j+1,i)
\end{equation}
with $p(i,j)=p(d=0 \vert x=i,y=j)$ ($0 \leq i,j \leq k-1$) satisfies $\vert Det(W_k) \vert=0$ for \textit{any} classical system of dimension $\leq k$. In WDCE we have $k=2$, since there are two possible experimental arrangements and we aim to test a classical model of dimension $2$. The matrix of interest is given by
\begin{equation} \label{W2}
W_2= 
 \begin{pmatrix}
  p(0,0)-p(1,0) & p(2,0)-p(3,0) \\
  p(0,1)-p(1,1) & p(2,1)-p(3,1) 
 \end{pmatrix},
\end{equation}
for which it can easily be verified that $\vert Det(W_2) \vert=0$ for the statistics predicted by quantum theory. Thus the experiment cannot rule out a classical explanation such as the hidden variable model we described earlier. Strikingly, as we show next, by slightly modifying Wheeler's scenario we can generate statistics that violate this dimensional witness and thus prove in a device-independent manner the incompatibility of a non-retrocausal hidden variable model with the generated data.
\par
\begin{figure}[t]
\centering\includegraphics[width=0.9\linewidth]{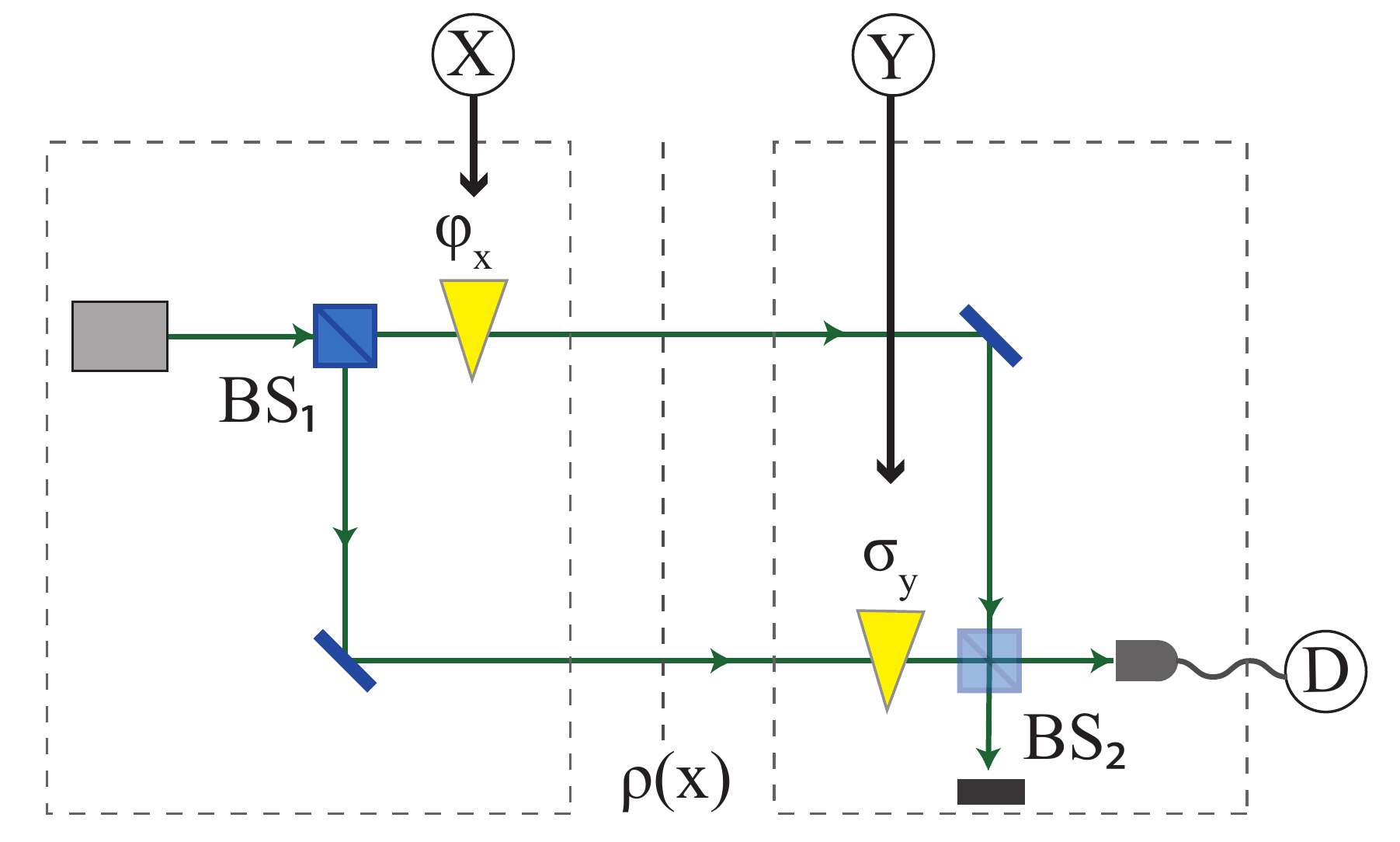}
\caption{Our proposed modification of WDCE, which can be used to discard two-dimensional HV models. The gray dashed lines allude to Fig.\ref{DAGfigures}(c). In addition to a phase shift $\phi_x$ in the preparation stage, another phase shift $\sigma_y$ is applied \textit{after} the single photon has entered the interferometer, for instance by slightly moving $\mathrm{BS}_2$.}
\label{ourDC}
\end{figure}
\par
In our proposed experiment, the interferometer is always closed. Instead of removing $\mathrm{BS}_2$, we gently displace it such that different measurement choices $y$ now correspond to a new phase shift $\sigma_y$ applied just before $\mathrm{BS}_2$ (Fig.\ref{ourDC}). From an experimental perspective, absorbing $\sigma_y$ into the preparation phase shift $\phi_x$ has no effect on the photon counting rates. However, to ensure that no retrocausality implies the independence of $Y$ and $\Lambda$, it is essential that $Y$ lies in the causal future of $\Lambda$ and hence of $X$. Therefore the choice of phase $\sigma_y$ must be delayed until long after the preparation phase $\phi_x$, which is applied shortly after the photon passes $\mathrm{BS}_1$ .

By considering this extra phase $\sigma_y$ and allowing for losses inside the interferometer, the statistics is given by
$p(x,y)=\frac{1}{4}\left(T_a^2+T_b^2\right)+\frac{T_aT_b}{2}\cos(\phi_x-\sigma_y)$,
where $0< T_a\leq 1$ and $0<T_b\leq 1$ are the real transmittance coefficients of each arm. In this case,
\begin{eqnarray}
\textrm{Det}(W_2) & &  = \frac{T_a^2T_b^2}{2}  \sin\left(\sigma_0-\sigma_1 \right) \sin\left(\frac{\phi_2}{2}-\frac{\phi_3}{2} \right)\\ \nonumber
 & & \left[ -\cos\left(\phi_0-\frac{\phi_2}{2}-\frac{\phi_2}{3} \right)+\cos\left(\phi_1-\frac{\phi_2}{2}-\frac{\phi_3}{2} \right)\right]
\end{eqnarray}

Choosing $\phi_0=0$, $\phi_1=\pi$, $\phi_2=-\pi/2$, $\phi_3=\pi/2$ we obtain that $\vert Det(W_2) \vert=T_a^2T_b^2 \sin\left(\sigma_0-\sigma_1 \right)$. If we make $\sigma_0=\pi/2$ and $\sigma_1=0$,  we obtain $\vert Det(W_2) \vert=T_a^2T_b^2$, that is, the dimension witness is violated for any transmittance strictly larger than zero. This test is also resilient to detection inefficiencies. If our detector has efficiency $\eta$ then $p^{\eta}(b=0 \vert x,y)=(1-\eta)+\eta p^{\eta=1}(b=0\vert x,y)$. Inserting this into $W_2$ we see that $\vert Det(W^{\eta}_2) \vert=\eta\vert Det(W^{\eta=1}_2) \vert$. Even though the violation is less at lower efficiencies $\eta$, it can in principle be observed for any positive efficiency.

So far we have also implicitly assumed that all noise terms are independent, and in particular that the hidden variable $\Lambda$ is independent of any noise term that might also influence the output of the interferometer. However, such a dependence would not be forbidden by causality, since we could imagine that the output of the interferometer depends on a noise variable $\Gamma$ that exists prior to the preparation of $\phi_x$ and $\Lambda$, and therefore might affect $\Lambda$ as well. One might wonder whether it is possible to rule out hidden variable models that allow this dependence. Surprisingly, it can be done. To achieve this we employ the dimension witness inequality \cite{Gallego2010}
\begin{equation}
\label{DWlinear}
\mathrm{I_{\mathrm{DW}}}=\mean{D_{00}}+\mean{D_{01}}+\mean{D_{10}}-\mean{D_{11}}-\mean{D_{20}} \leq 3,
\end{equation}
where $\mean{D_{xy}}=p(d=0\vert x,y)-p(d=1\vert x,y)$. This inequality involves $3$ preparations and $2$ measurements and its violation witnesses incompatibility with any HV model of dimension $2$, even in the presence of correlations between the preparation and measurement devices, i.e. even when both $\Lambda$ and the outcome of detector $D$ have access to shared information.
Using the same setup as above, we obtain $\mean{D}_{xy}= T_aT_b\cos{(\phi_x-\sigma_y)} $. For $T_a=T_b=1$, and choosing $\sigma_0=\pi/2;\; \sigma_1=0;\; \phi_0=\pi /4;\;\phi_1=3\pi /4;\; \phi_2=-\pi/2$ we obtain the optimum quantum violation given by $\mathrm{I}_{\mathrm{Q}}=1+2\sqrt{2}$ \cite{Ahrens2012}. 

\emph{Quantifying Retrocausality --} Altogether, our proposed modification of WDCE can exclude in a DI manner all non-retrocausal classical models with dimension $d=2$. Conversely, if one allows retrocausality (see Fig. \ref{DAGfigures}(b)) any observed distribution $p(d \vert x,y)$ could be simulated classically. But how much retrocausality is actually needed to reproduce the quantum predictions? To answer this we need to quantify the strength of the causal influence $Y \rightarrow \Lambda$ in the DAG. Without this causal arrow, and allowing for correlations between preparation and measurement devices (described by a HV $\Gamma$), $p(\lambda\vert x,\gamma,y)=p(\lambda\vert x,\gamma,y^{\prime})$. This leads us to consider a measure of retrocausality given by
\begin{equation}
\label{retro}
\mathrm{R}_{Y \rightarrow \Lambda}= \sup_{\lambda,x,y,y^{\prime}}  \sum_{\gamma}p(\gamma)\vert p(\lambda\vert x,\gamma,y) -p(\lambda\vert x,\gamma,y^{\prime}) \vert,
\end{equation}
the maximum shift in the probability of the HV $\Lambda$ produced by changes in the measurement setting $Y$ for a preparation $X$ (averaged over $\Gamma$ since we do not have empirical access to it). As explained in the Appendix, given an observed value of the dimension witness $\mathrm{I_{\mathrm{DW}}}$, the minimum value of $\mathrm{R}_{Y \rightarrow \Lambda}$ required to explain it is
\begin{equation}
\min \mathrm{R}_{Y \rightarrow \Lambda}= \max{\left[\frac{\mathrm{I}-3}{4},0 \right]},
\end{equation}
thus showing that the maximal possible value of $\mathrm{I}_{\mathrm{Q}}=1+2\sqrt{2}$ allows one to exclude any retrocausal model with $\mathrm{R}_{Y \rightarrow \Lambda} \lesssim 0.207$.

\emph{Discussion--} The vague notion of wave-particle objectivity has been analyzed in different ways by different authors. For instance, Wheeler \cite{Wheeler1978,Wheeler1984} associated wave and particle notions to the possibility (resp. impossibility) of a photon being in a path superposition inside the interferometer. In turn, the IT model \cite{Ionicioiu2011} associates wave and particle notions to the statistics obtained at the detectors, implying the label `wave' if the statistics depend on the phase shift $\phi$, and `particle' otherwise. This interpretation may be criticised for implying that $\lambda=$`wave' cannot produce statistics independent of $\phi$, whereas in an open interferometer this is precisely what we would expect from a classical wave. Furthermore, it implies a correlation between $\Lambda$ and the detector $D$ which, as we showed using causal modeling, makes it trivially incompatible with the quantum predictions and no retrocausality.

To avoid the difficulties arising from particular interpretations of ``wave-particle objectivity", one could consider different ways of describing delayed-choice experiments. In Ref.\cite{Rossi2017}, it was shown that a simple argument based on causal models could provide conceptual insights into entanglement-assisted delayed choice experiments (see Appendix \ref{entqdce}). We have shown that causal models can also shed light on the simpler experiments proposed by Wheeler \cite{Wheeler1978,Wheeler1984}, and the quantum variants proposed in Refs. \cite{Ionicioiu2011,Auccaise2012}.

We argued that, regardless of any categorization as a wave or particle, \emph{any} pre-defined classical state of the photon that is supposed to reproduce the results of WDCE or QDCE should have two values, corresponding to the dimension of the quantum system being probed. This holds for both Wheeler's and the IT model's conception of wave-particle objectivity, and indeed both can be cast as two-dimensional classical hidden variable models. By taking the dimensionality to be the main relevant feature, we showed that these models could be excluded using DI methods, provided the experiment is modified such that $\mathrm{BS}_2$ is always present. Since 'wave-like' behaviour (interference) is always present in our experiment, our approach does not rely on making any kind of wave/particle distinction: we merely probe how much information must be conveyed by a classical variable in order to produce the observed interference. In this respect, our proposal is conceptually distinct from previous approaches. Moreover, our results extend to any two-dimensional hidden variable, whether it be a 'wave', a 'particle' or something queer in-between.

This gives us a new perspective on what is counterintuitive in delayed choice experiments: the fact that any classical explanation requires a variable with more dimensions than its quantum counterpart, or else requires retrocausal influences. Our results also show the benefits of viewing quantum phenomena from a causal perspective. For applications, since the PAM scenario plays an important role in recent work on quantum key distribution \cite{Pawlowski2011,Lunghi2015,Brask2017}, our work suggests that interesting quantum information protocols could be performed with setups as simple as a Mach-Zehnder interferometer.

\begin{acknowledgments}
We acknowledge financial support from the Brazilian ministries MEC and MCTIC. We would like to thank Romeu Rossi for discussions that inspired this investigation. 
\end{acknowledgments}


\begin{thebibliography}{30}
\bibitem [{not({\natexlab{a}})}]{note1}%
  \BibitemOpen
   {It is essential that a
  single photon source is used, such that both detectors never \textit{click}
  simultaneously. This guarantees that each photon cannot be modeled as a
  classical wave that is quantized only at the detector
  \cite{Paul1982}.}\BibitemShut {Stop}%
\bibitem [{not({\natexlab{b}})}]{note3}%
  \BibitemOpen
  {These local noise
  variables are usually left implicit in the DAG.}\BibitemShut {Stop}%
\bibitem [{not({\natexlab{c}})}]{note4}%
  \BibitemOpen
  {Such a dependence of
  $\lambda$ on $Y$ would constitute a retrocausal influence, represented by a
  causal arrow $Y \rightarrow \lambda$.}\BibitemShut {Stop}%
\bibitem [{not({\natexlab{e}})}]{note5}%
  \BibitemOpen
  {Since we assume that a
  single photon enter in the interferometer only one of the detectors will
  click in each experimental run, that is, $p(d=1)=1-p(e=1)$, thus rendering
  superfluous the analysis of both detectors.}\BibitemShut {Stop}%
\end{thebibliography}

%

\appendix

\section{A HV model for the quantum delayed choice experiment}
In the main text we have shown that a simple HV model (with no retrocausality) can reproduce all the statistical predictions of quantum mechanics in the usual delayed choice experiment. This causal model admits an interpretation as a hidden variable model as follows. Imagine that the photon is indeed a particle, and the values $\lambda=\{0,1 \}$ correspond to the particle taking the upper or lower path, respectively, through the interferometer. The choice $X$ of phase may affect the probability $p(\lambda|x)$ of the particle's location. Although this state cannot depend in advance on $Y$, it is reasonable to allow that the hidden variable reacts to the presence or absence of BS2 on arrival. If BS2 is present, it continues on its way -- but if BS2 is absent, it is replaced by the 'local noise' $u_D$ that triggers detector $D$ or $E$ with equal likelihood. This model violates the extra assumption of Refs. \cite{Ionicioiu2011,Ionicioiu2015} because the hidden variable state $p(\lambda|x)$ does not by itself determine whether interference is observed or not. Nevertheless, we argue that this model is wave-particle objective because it describes a 'particle' that is located only in one of the interferometer paths throughout the experiment.
\par\begin{figure}[t]
\centering\includegraphics[width=0.8\linewidth]{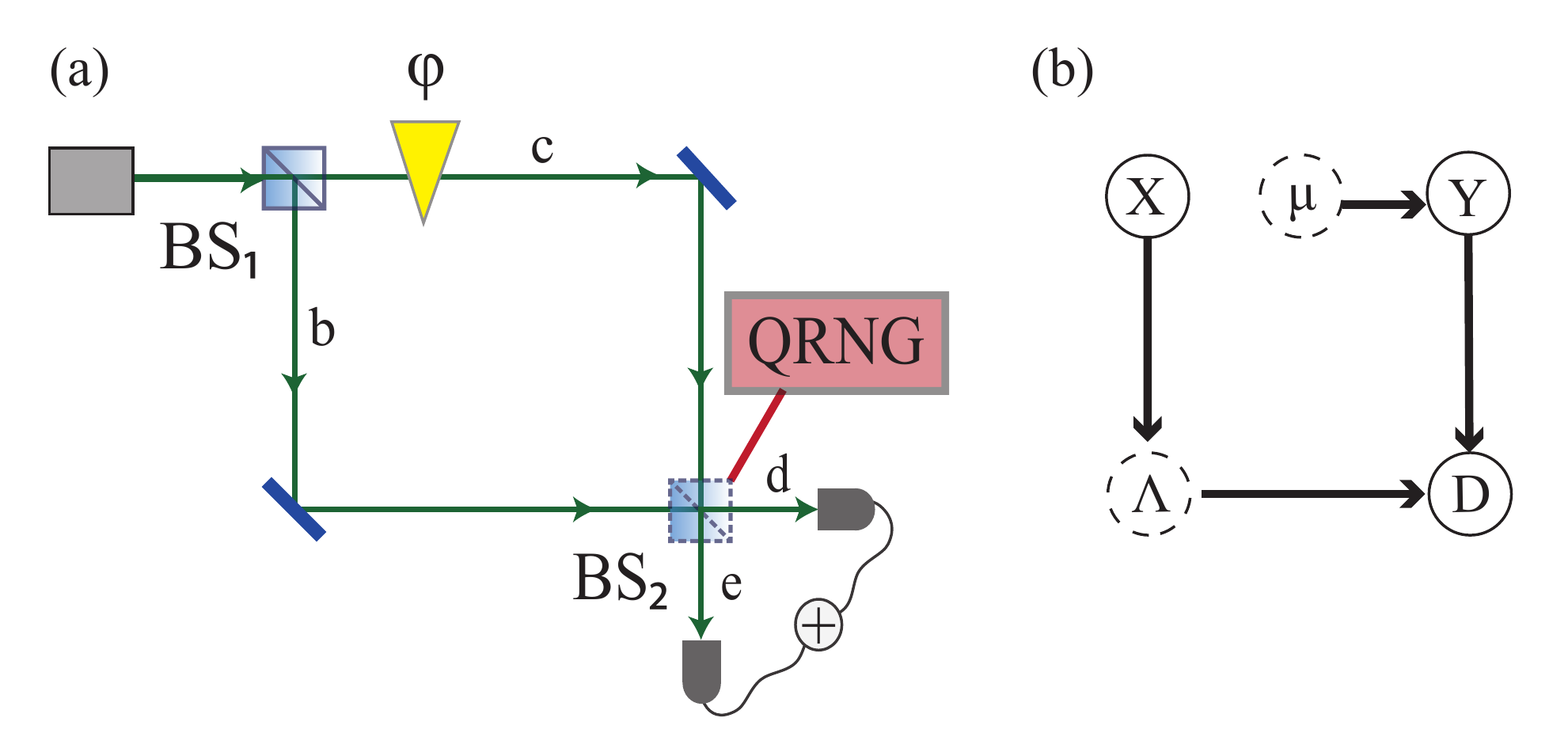}
\caption{(a)In the QDCE, the choice of removing or not the beam-splitter $\mathrm{BS}_2$ in a Mach-Zehnder interferometer is controlled by a quantum random number generator (QRNG), \emph{i.e.} a quantum system in a coherent superposition. (b) In the causal analysis of the QDCE we replace the quantum control by another hidden variable that is independent of all variables but $Y$. }
\label{supmatfig}
\end{figure}
\par
Building up on WDCE, Ref. \cite{Ionicioiu2011} has proposed to use a quantum system as the control of which experiment to perform (Fig.\ref{supmatfig}). More precisely, starting with a quantum control in a coherent superposition $\cos{\alpha} \ket{0} +\sin{\alpha} \ket{1}$ such that if the control is $\ket{0}$/$\ket{1}$ the $\mathrm{BS}_2$ is not present/present we have that just before the detectors the joint state system/control is given by $\cos{\alpha} \ket{0}\ket{\mathrm{particle}} +\sin{\alpha} \ket{1}\ket{\mathrm{wave}}$ with $\ket{\mathrm{particle}}=(1/\sqrt{2})(\ket{0}+e^{i \phi}\ket{1})$ and $\ket{\mathrm{wave}}=e^{i\phi /2}(\cos{\frac{\phi}{2}}\ket{0}-i\sin{\frac{\phi}{2}}\ket{1})$. As opposed to the case with a classical control (defined by the experimenter), in this case the control is determined by the measurement of the quantum system in the basis $\ket{0}, \ket{1}$. For that reason, instead of describing the experiment in terms of $p(d \vert x,y)$ we use instead $p(d,y\vert x)$. In other terms, in the case of a classical control the distribution $p(y)$ does not carry any relevant information (since it is chosen from the experimenter) while in the case of a quantum control we have to explicitly take that into account when trying to construct a classical causal model. A classical description of the quantum delayed choice experiment implies that the observed distribution should be decomposed as
\begin{equation}
\label{HVquantum}
p(d,y \vert x)= \sum_{\lambda,\mu} p(d \vert y,\lambda)p(\lambda \vert x)p(y \vert \mu) p(\mu).
\end{equation}
That is, we replace the quantum control by another hidden variable governed by a probability distribution $p(\mu)$ that is independent of all variables but $Y$. 

In the experiment with a quantum control the observed distribution is given by
\begin{eqnarray}
p(d,y\vert x)  = & & (1/2) \delta_{y,0}\cos^2{\alpha}+ \\ \nonumber & &(1/2)\delta_{y,1}\sin^2{\alpha}({\delta_{d,0}}\cos^2{\frac{\phi}{2}}+{\delta_{d,1}}\sin^2{\frac{\phi}{2}}).
\end{eqnarray}
To reproduce this statistics with the classical model \eqref{HVquantum} we can employ the same classical causal model introduced in the main text with the only difference that now we have to choose $p(\lambda \vert \mu)=\delta_{y,\mu}$ with $p(\mu=0)=\cos^2{\alpha}$ and $p(\mu=1)=\sin^2{\alpha}$.

\section{The causal analysis of QDCEs with and without entanglement assistance.}\label{entqdce}
In \cite{Rossi2017} causal analysis was elegantly applied to entanglement-assisted quantum delayed choice experiments. The entanglement-assisted protocol extends the proposal of Ref. \cite{Ionicioiu2011} by having the control system not merely in a superposition, but \textit{entangled} to another system that is measured at space-like separation from the output of the interferometer. In such a set-up, Ref. \cite{Rossi2017} argued that there are two independent hidden variables, and a straightforward causal analysis revealed that no classical model with two 
hidden variables could account for the statistical independencies predicted by the experiment. It follows that any attempt to extend the IT model to the entanglement-assisted QDCE would be ruled out by this method. 

This result led us to wonder whether causal analysis could be used to rule out classical explanations in the much simpler set-ups that we consider, which don't rely on having entanglement assistance. Indeed, we concluded that such experiments cannot be used to rule out classical models in a device-independent way. We therefore proposed a modified version of the experiment that would be sufficient, but still without relying on entanglement assistance. By contrast, the no-go result of \cite{Rossi2017} depends upon the control system being entangled to an auxiliary system. The two approaches are therefore complementary: whereas the result of \cite{Rossi2017} requires entanglement assistance but does not assume anything about the dimension of the classical system, our approach requires no entanglement assistance, but does assume a restriction on the classical variable's dimension.

\section{Quantifying retrocausality}
In a retrocausal model we allow the state of the system $\Lambda$ at an earlier time to depend on the later choice $Y$ of which kind of measurement to perform. In this case, any observed probability distribution compatible with such model decomposes as
\begin{equation}
p(d \vert x,y) =\sum_{\lambda} p(d \vert y,\lambda) p(\lambda \vert x,y).
\end{equation}
Clearly, if the random variables $D$ and $\Lambda$ have the same dimension (or $\Lambda$ has a higher dimension) one can generate any distribution $p(d \vert x,y)$ since we can simply make $p(\lambda \vert x,y)=p(d \vert x,y)$ and $ p(d \vert y,\lambda)=\delta_{d,\lambda}$. That makes important to not only allow for such retrocausal influence but also to quantify how much of it is actually necessary to reproduce a given observation.

Different measures of retrocausality can be considered. At first, since in a non-retrocausal model we have that $p(\lambda,y)=p(\lambda)p(y)$, one might consider natural to consider as a measure the trace distance between $\mathrm{TD}=(1/2)\sum_{y,\lambda}\vert p(\lambda,y)-p(\lambda)p(y) \vert$. Interestingly, however, this measure can be zero even thought the dimension witness \eqref{DWlinear} is maximally violated (with a retrocausal HV model of dimensionality $2$). To see this consider a retrocausal model where the value of $\Lambda$ is a deterministic function of $X$ and $Y$ such that $\lambda=0$ if $x=0$, $\lambda=y$ if $x=1$ and $\lambda=y \oplus 1$ if $x=2$ and where $D$ is a deterministic function of $\Lambda$ such that $d=\lambda$. This retrocausal model achieves $\mathrm{I}_{\mathrm{DW}}=5$ using a HV of dimension 2. Nevertheless we see that $\mathrm{TD}=0$ since $p(\lambda \vert y)= p(\lambda \vert y^{\prime}))$.

While the causal constraint $p(\lambda,y)=p(\lambda)p(y)$ is indeed a consequence of non-retrocausal models, the causal constraint imposed by the Markov decomposition associated with the DAG is actually given by $p(\lambda \vert x,y)=p(\lambda \vert x,y^{\prime})$ (in other terms $p(\lambda \vert x,y)=p(\lambda \vert x)$). This considerations naturally lead us to consider the measure of retrocausality introduced in the main text \eqref{retro} given by 
\begin{equation}
\mathrm{R}_{Y \rightarrow \Lambda}= \sup_{\lambda,x,y,y^{\prime}} \sum_{\gamma} p(\gamma) \vert p(\lambda\vert x,\gamma,y) -p(\lambda\vert x,\gamma,y^{\prime}) \vert.
\end{equation}
Given a certain value of the dimension witness $\mathrm{I}_{\mathrm{DW}}$ we are interested to know what the the minimum value of $\mathrm{R}_{Y \rightarrow \Lambda}$ required to explain it with a retrocausal HV model. Following an approach similar to those in Refs. \cite{Chaves2015b,Brito2017} one can cast this minimization problem as a linear program that can be easily solved to show that $\min \mathrm{R}_{Y \rightarrow \Lambda}= \max{\left[\frac{\mathrm{I}-3}{4},0 \right]}$.

Interestingly, this measure is not maximal even for the maximal violation of the dimension witness inequality. Namely, we have $\mathrm{R}_{Y \rightarrow \Lambda}=1/2$ when $\mathrm{I}_{\mathrm{DW}}=5$ (maximal violation). To understand that, notice that the following two deterministic strategies lead to $\mathrm{I}_{\mathrm{DW}}=5$: i)  $\lambda=0$ if $x=0$, $\lambda=y$ if $x=1$ and $\lambda=y \oplus 1$ if $x=2$ and where $D$ is a deterministic function of $\Lambda$ such that $d=\lambda$ and ii)  $\lambda=y$ if $x=0$, $\lambda=0$ if $x=1$ and $\lambda=1$ if $x=2$ and where $D$ is a deterministic function of $\Lambda$ and $Y$ such that $d=\lambda \oplus y$. We can express the deterministic function $\lambda=f(x,y)$ as a vector $(f(0,0),f(0,1),f(1,0),f(1,1),f(2,0),f(2,1))$. For the first deterministic strategy above the vector is given by $(0,0,0,1,1,0)$ and for the second it is given by $(0,1,0,0,1,1)$. Clearly, for each of these strategies we have that $\mathrm{R}_{Y \rightarrow \Lambda}=1$. For instance, for the first strategy $p(\lambda=0 \vert x=1,y=0)=1$ while $p(\lambda=0 \vert x=1,y=1)=0$. Nonetheless, if we probabilistically choose between both deterministic strategies we reduce $\mathrm{R}_{Y \rightarrow \Lambda}$ since, for example, for the second strategy we have that $p(\lambda=0 \vert x=1,y=0)=1$.

\end{document}